\documentstyle{mn}	

\newif\ifAMStwofonts

\def\refit{}\def\refbf{}
\def\myref#1,#2,#3.{{\refit #1\/}, {\refbf #2}, #3\par\noindent}
\def\aa,#1,#2.{\myref{A\&A},#1,#2.}
\def\acta,#1,#2.{\myref{Acta Astron.},#1,#2.}
\def\annrev,#1,#2.{\myref{ARAA},#1,#2.}
\def\aj,#1,#2.{\myref{AJ},#1,#2.}
\def\apj,#1,#2.{\myref{ApJ},#1,#2.}
\def\apjsupp,#1,#2.{\myref{ApJS},#1,#2.}
\def\apspsci,#1,#2.{\myref{Ap\&SS},#1,#2.}
\def\aasupp,#1,#2.{\myref{AA\ Supp.},#1,#2.}
\def\ica,#1,#2.{\myref{Icarus},#1,#2.}
\def\grg,#1,#2.{\myref{GRG},#1,#2.}
\def\jaa,#1,#2.{\myref{J.\ Astr.\ Astrophys.},#1,#2.}
\def\mnras,#1,#2.{\myref{MNRAS},#1,#2.}
\def\nat,#1,#2.{\myref{Nature},#1,#2.}
\def\pasp,#1,#2.{\myref{PASP},#1,#2.}
\def\pasj,#1,#2.{\myref{PASJ},#1,#2.}
\def\physrev,#1,#2.{\myref{Phys.\ Rev.},#1,#2.}
\def\physrevlett,#1,#2.{\myref{Phys.\ Rev.\ Lett.},#1,#2.}
\def\physrevD,#1,#2.{\myref{Phys.\ Rev.\ D},#1,#2.}
\def\procroy,#1,#2.{\myref{Proc.\ Roy.\ Soc.},#1,#2.}
\def\revmod,#1,#2.{\myref{Rev.\ Mod.\ Phys.},#1,#2.}
\def\sova,#1,#2.{\myref{SvA},#1,#2.}

\def\etal{{\it et\thinspace al\/}}

\def\ie{{i.e}}
\def\eg{{e.g}}

\def\cf{{cf}}
\def\kms{\mbox{km s$^{-1}$}}
\def\mpc{\mbox{Mpc}}

\def\halff{{\textstyle{1\over2}}}
\def\frac(#1/#2){{\textstyle{#1\over#2}}}

\edef\vbar{|}
\def\subrm#1{_{\rm #1}}
\catcode`|=\active \let|=\subrm

\def\mod#1{\left\vbar#1\right\vbar} 
 
\def\pd#1#2{{\upartial#1\over \upartial#2}} \def\lr#1{\left(#1\right)}

\def\const{\hbox{constant}}

\newcount\psw 
\def\computelabs{
  \psw=\number\epsfxsize \divide\psw by 65536 
}%
\newbox\labelbox
\newdimen\yy\newdimen\xx
\newdimen\hair\hair=3pt
\def\setlabel#1#2#3{%
\setbox\labelbox\hbox{$#1$}%
\vbox to 0pt{\yy=#3 pt 
\divide \yy by \mag%
\multiply \yy by 1010%
\multiply \yy by \the\psw%
\advance\yy by 6pt 
\kern-\yy%
\hbox to 0pt{%
\xx=#2 pt 
\divide \xx by \mag%
\multiply \xx by 1010%
\multiply \xx by \psw
\kern\xx\box\labelbox\hss}\vss}\ifvmode\nointerlineskip\fi}

\def\reference{\item}
\def\vmax{v|{m}}
\def\rd{R|d}
\def\rh{R|h}
\def\MI{{\cal M}_I}
\def\epsl{\epsilon_{l}}
\def\epsd{\epsilon|{d}}
\def\epscr{\epsilon|{c}}
\def\epsm{\epsilon|{m}}
\def\m2l{\Upsilon_I}

\def\Msun{{\ifmmode M_{\sun} \else $M_{\sun}$ \fi}}
\def\msun{{\ifmmode m_{\sun} \else $m_{\sun}$ \fi}}
\def\Lsun{{\ifmmode L_{\sun} \else $L_{\sun}$ \fi}}
\def\rsun{{\ifmmode r_{\sun} \else $r_{\sun}$ \fi}}
\def\Rsun{{\ifmmode R_{\sun} \else $R_{\sun}$ \fi}}
\def\pc{{\ifmmode \hbox{pc} \else {pc} \fi}}

\def\ELNr{Efstathiou, Lake \& Negroponte (1982)}
\def\ELN{ELN}

\input epsf

\ifoldfss
  \ifCUPmtlplainloaded \else
    \NewTextAlphabet{textbfit} {cmbxti10} {}
    \NewTextAlphabet{textbfss} {cmssbx10} {}
    \NewMathAlphabet{mathbfit} {cmbxti10} {} 
    \NewMathAlphabet{mathbfss} {cmssbx10} {} 
  \fi
  \ifAMStwofonts
    \ifCUPmtlplainloaded \else
      \NewSymbolFont{upmath} {eurm10}
      \NewSymbolFont{AMSa} {msam10}
      \NewMathSymbol{\upi}     {0}{upmath}{19}
      \NewMathSymbol{\umu}     {0}{upmath}{16}
      \NewMathSymbol{\upartial}{0}{upmath}{40}
      \NewMathSymbol{\leqslant}{3}{AMSa}{36}
      \NewMathSymbol{\geqslant}{3}{AMSa}{3E}

      \let\leq=\leqslant 
       
    \fi
  \fi
\fi 

\ifnfssone
  \newmathalphabet{\mathit}
  \addtoversion{normal}{\mathit}{cmr}{m}{it}
  \addtoversion{bold}{\mathit}{cmr}{bx}{it}
  \newmathalphabet{\mathbfit} 
  \addtoversion{normal}{\mathbfit}{cmr}{bx}{it}
  \addtoversion{bold}{\mathbfit}{cmr}{bx}{it}
  \newmathalphabet{\mathbfss} 
  \addtoversion{normal}{\mathbfss}{cmss}{bx}{n}
  \addtoversion{bold}{\mathbfss}{cmss}{bx}{n}
  \ifAMStwofonts
    \ifCUPmtlplainloaded \else
      \UseAMStwoboldmath
      \makeatletter
      \new@mathgroup\upmath@group
      \define@mathgroup\mv@normal\upmath@group{eur}{m}{n}
      \define@mathgroup\mv@bold\upmath@group{eur}{b}{n}
      \edef\UPM{\hexnumber\upmath@group}
      \new@mathgroup\amsa@group
      \define@mathgroup\mv@normal\amsa@group{msa}{m}{n}
      \define@mathgroup\mv@bold\amsa@group{msa}{m}{n}
      \edef\AMSa{\hexnumber\amsa@group}
      \makeatother
      \mathchardef\upi="0\UPM19
      \mathchardef\umu="0\UPM16
      \mathchardef\upartial="0\UPM40
      \mathchardef\leqslant="3\AMSa36
      \mathchardef\geqslant="3\AMSa3E

      \let\leq=\leqslant 

    \fi
  \fi
\fi 

\ifnfsstwo
  \DeclareMathAlphabet{\mathbfit}{OT1}{cmr}{bx}{it}
  \SetMathAlphabet\mathbfit{bold}{OT1}{cmr}{bx}{it}
  \DeclareMathAlphabet{\mathbfss}{OT1}{cmss}{bx}{n}
  \SetMathAlphabet\mathbfss{bold}{OT1}{cmss}{bx}{n}
  \ifAMStwofonts
    \ifCUPmtlplainloaded \else
      \DeclareSymbolFont{UPM}{U}{eur}{m}{n}
      \SetSymbolFont{UPM}{bold}{U}{eur}{b}{n}
      \DeclareSymbolFont{AMSa}{U}{msa}{m}{n}
      \DeclareMathSymbol{\upi}{0}{UPM}{"19}
      \DeclareMathSymbol{\umu}{0}{UPM}{"16}
      \DeclareMathSymbol{\upartial}{0}{UPM}{"40}
      \DeclareMathSymbol{\leqslant}{3}{AMSa}{"36}
      \DeclareMathSymbol{\geqslant}{3}{AMSa}{"3E}

      \let\leq=\leqslant 

    \fi
  \fi
\fi 

\ifCUPmtlplainloaded \else
  \ifAMStwofonts \else 
    \def\upi{\pi}
    \def\umu{\mu}
    \def\upartial{\partial}
  \fi
\fi

\title[] 
{Global stability and the mass-to-light ratio of
galactic disks}
\author[]
{D. Syer, Shude Mao, and H.J. Mo
\thanks{E-mail: (syer, smao, hom)@mpa-garching.mpg.de} \\
	Max-Planck-Institut f\"ur Astrophysik
	Karl-Schwarzschild-Strasse 1, 85748 Garching, Germany}
\date{Accepted ........
      Received .......;
      in original form .......}

\pagerange{\pageref{firstpage}--\pageref{lastpage}}
\pubyear{1997}

\begin{document}
\maketitle
\label{firstpage}

\begin{abstract}
We examine the global stability of an exponential stellar disk
embedded in a dark matter halo and constrain the mass-to-light ratio
of observed disks in the I-band, $\m2l$.  Assuming only that the
radial surface density distribution of disks is exponential, we derive
an analytic upper limit: $\m2l\la 2.4h$, for a Hubble constant of
$100h~\kms \mpc^{-1}$.  Using $N$-body simulations we derive a
stability criterion significantly different from that of previous
authors.  Using this criterion we argue that, almost independent of
the concentration of the halo, $\m2l\la 1.9h$.  We discuss this result
in relation to other independent determinations of $\m2l$, its
limitations and its implications for theories of disk formation and
barred galaxies.
\end{abstract}


\section{Introduction}
Stellar disks are a potentially powerful probe of the mass
distribution in spiral galaxies.  In particular, the rotation curve
(defined in Section \ref{expsec}) is a direct probe of the
gravitational force in the disk.  The fact that the rotation curves of
spiral galaxies are rather flat is usually taken to imply the presence
of an extended halo of dark matter (\eg{.} Freeman 1970, Persic \& Salucci
1991).  Dark halos have been traditionally modelled by isothermal
spheres with homogeneous cores, and assigned a mass-to-light ratio
according to the `maximum-disk' hypothesis, in which the largest mass
possible is assigned to the disk consistent with the rotation curve
(Carignan \& Freeman 1985).  Recent theoretical work has suggested
that the isothermal form of the halo may not be appropriate (Navarro,
Frenk \& White 1996, Moore \etal{.}  1997, Kravstov \etal{.} 1997).
Rix \& Courteau
(1997) have recently argued that the Tully-Fisher relation (Tully \&
Fisher 1977) has too little scatter to be consistent with the
maximum-disk hypothesis.

Dalcanton, Spergel \& Summers (1997), and Mo, Mao \& White (1997)
point up the need for a consistent and accurate picture of disk
mass-to-light ratios for the theory of disk formation, and to
understand the Tully-Fisher relation.  As was pointed out by Rix \&
Courteau (1997), the maximum disk hypothesis and the Tully-Fisher
relation are together incompatible with a near universal mass-to-light
ratio, $\Upsilon$.  Yet, the stellar populations of spiral disks at
$z=0$ are generally believed to have rather uniform properties.  In
particular they all have similar colours, which leads one to believe
that they should have similar $\Upsilon$.

Ostriker \& Peebles (1973) argued that extended halos of dark matter
in disk galaxies are required to stabilize the disks against global
bar instabilities.  They were able to draw very strong conclusions
from $N$-body simulations with only 300 particles.  \ELNr{} (\ELN)
conducted larger $N$-body experiments specifically with disk
components appropriate for real galaxies, and concluded that
$\Upsilon|{bol}\leq1.5\pm0.2$. 

\ELN{} compared their numerical results with only 12 galaxies due to
limited data at that time. Since then, observations of spiral galaxies
have improved tremendously owing to the great interest in the
Tully-Fisher relation as a distance indicator (Giovanelli \etal{.}
1997 and references therein).  Large samples of spiral galaxies are
now available (\eg{.} Mathewson \& Ford 1996; Courteau 1996, 1997). In
this paper, we re-examine the constraints from considerations of
global stability which may be placed on the mass-to-light ratio of
disks.

In the next section we review the properties of exponential disks, and
their relation to real disk galaxies.  In Section \ref{stabsec} we
review the physics of global instability in disks.  In Section
\ref{obsec} we describe the observations of disk galaxies, and how
they can be used to constrain $\Upsilon$.  In Section \ref{numsec} we
describe our $N$-body simulations.  In Section \ref{detsec} we discuss
the results of independent determinations of mass-to-light ratios of
galactic disks.  In Section \ref{dissec} we discuss the implications
of our results and draw conclusions.

\section{Exponential disks}\label{expsec}
The luminous disks of spiral galaxies are commonly modelled by an
exponential surface brightness distribution:
\begin{equation}
\mu(R) = {L|d\over2\pi\rd^2} \exp(-R/\rd)
\label{muexp}
\end{equation}
where $R$ is the usual cylindrical radius, and $L|d$ is the total
luminosity of the disk.  Here we collect some notation and a number of
useful results relating to exponential disks.

The disk has a mass $M|d$, and a mass-to-light ratio in the $I$-band
$\m2l$ in solar units.  Thus the surface mass density of the disk is
\begin{equation}
\Sigma(R) = \mu(R)\m2l = {M|d\over2\pi\rd^2} \exp(-R/\rd).
\label{Sigexp}
\end{equation}
The gravitational potential in the disk $\Phi$
is conveniently decomposed into contributions from the disk and a halo:
\begin{equation}
\Phi = \Phi|{d} + \Phi|{h}.
\label{phid}
\end{equation}
(We use the subscripts `d' for `disk', and `h' for `halo' throughout.)
We assume for the present purposes that the halo is spherical, and
usually we think of it as being composed of dark matter, but it may
also contain a stellar component (\eg{.} the `bulge' of an earlier
type spiral).

The speed of test particles on circular orbits $v|c$ as a function of
$R$ is given by
\begin{equation}
v^2|c(R) = -\pd{\Phi}{R}.
\label{rotc}
\end{equation}
For the sake of definiteness, we shall refer to $v|c(R)$ as the `true
rotation curve' of the system (or just `the rotation curve').  The
observed line-of-sight velocity of a tracer population corrected for
inclination we refer to as the rotation curve of the tracer.  For
instance, the HI rotation curve (apart from small contributions from
turbulent motion) is thought to be a good measure of the true rotation
curve as long as the system is axisymmetric.  Some of the samples of
galaxies used in this paper have rotation curves measured in
H$_\alpha$, but analyzed in a way which is supposed to maximize the
agreement with HI (Raychaudhury \etal{} 1997, Courteau 1997).

Following \ELNr{} we define the dimensionless
quantity
\begin{equation}
\epsm = {\vmax\over(GM|d/\rd)^{1/2}},
\label{epsdef}
\end{equation}
where $\vmax$ is the maximum value of $v|c$, and $G$ is the
gravitational constant.  The rotation
curve of an isolated exponential disk ($\Phi|h=0$) is (Freeman 1970):
\begin{equation}
v^2|d(R) = {2 GM|d\over\rd} y^2[I_0(y)K_0(y) - I_1(y)K_1(y)],
\label{rotexp}
\end{equation}
where $y=R/(2\rd)$. $I_i$ and $K_i$ are modified Bessel functions.
An isolated disk has $\epsm\equiv\epsd\approx0.63$
and a disk embedded in a halo has $\epsm>\epsd$.

\section{Global stability}\label{stabsec}
An isolated thin exponential disk is known to be violently unstable,
as are all isolated thin disks.  The fastest growing instability is
generically global (the whole system is deformed) and bar shaped.  The
instability in conservative and/or collisionless systems is generally
purely dynamical (growth timescale of order the crossing time).
Secular instabilities set in in systems with dissipation and/or
collisions generally on longer timescales.  The disk may be stabilized
if the disk is thickened in the vertical direction, as exemplified by
the Maclaurin sequence of gaseous spheroids (Chandrasekhar 1969): near
spherical members of the family are stable; they become first
secularly and then dynamically unstable as the eccentricity is
increased (see for example Christodoulou, Schlosman \& Tohline 1995).
A similar pattern is found for the stellar analogues of the Maclaurin
spheroids.  Stability may also be conferred by adding an external
potential---in the extreme limit that the disk is composed of test
particles it is trivially stable.  As was pointed out by Ostriker \&
Peebles (1973) this mechanism can be used on rather general grounds to
argue that spiral galaxies contain halos of dark matter.

There has been some debate as whether a disk with a given eccentricity
and/or external potential is expected to be stable on theoretical
grounds.  Many attempts have been made to find a single parameter
which delineates a boundary between stability and instability.
Ostriker \& Peebles (1973) used as a parameter the ratio
\begin{equation}
t|{OP}=T|{mean}/W
\label{tOP}
\end{equation}
of the mean kinetic to total potential energy. The stellar disks
studied by Ostriker \& Peebles (1973) were stable if $t|{OP}<0.14$ to
very high accuracy.  Remarkably, this criterion was shown to unify
their results with all existing results on the stability of disks.
Subsequent authors pointed to various insufficiencies of $t|{OP}$ as
applied to specific cases (Miller 1978, Miller \& Smith
1979).  \ELNr{} refined the definition to patch up some of these
difficulties, defining a related parameter $t_*$, but in the end
concluded that $\epsm$ (cf. equation \ref{epsdef})
was a better indicator of stability for their
models.  In particular they found that all models with $\epsm\la1.1$
were unstable, whereas the stability boundary in $t_*$ depended on the
halo concentration.  Christodoulou, Schlosman \& Tohline (1995)
proposed a further refinement of $t_*$ by analogy with the Maclaurin
series, defining a new parameter
\begin{equation}
\alpha=\sqrt{\halff f t},
\end{equation}
where $t=T/W$ and $f$ is a form factor dependent on the shape of the
disk.  Both $t$ and $f$ were given precise definitions only for the
Maclaurin sequences.  They showed that $\alpha$ led to a stability
criterion which was broadly consistent with that of \ELN{}.  However,
$\alpha$ was not defined properly for disks with external potentials,
and they failed to address the issue of the halo concentration.

The importance of global stability is this: the disks in spiral
galaxies appear to be long lived, and the majority do not contain
bars.  In the next section we describe how the apparent stability of
galactic disks can be used to constrain their mass-to-light ratio.

\begin{figure}
\epsfxsize=.9\hsize
\computelabs
\setlabel{\epsl h^{-\halff}}{-.05}{-.45}
\setlabel{\mu_0 (I-\hbox{mag/sec}^2)}{.53}{-.94}
\centerline{\epsfnormal{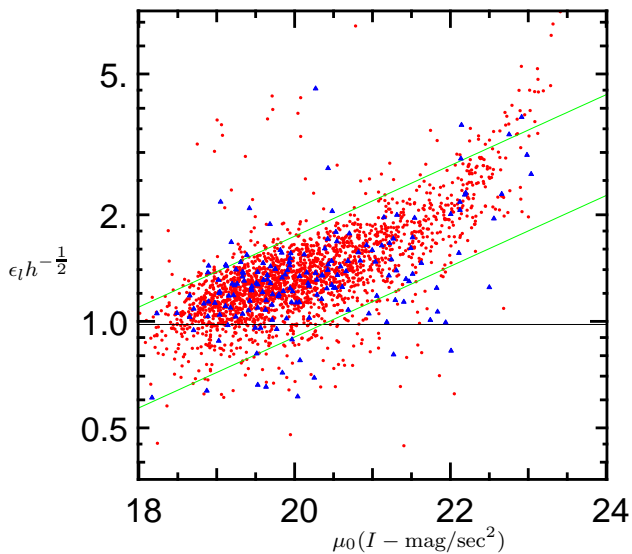}}
\vskip\baselineskip
\caption{ The quantity $\epsl h^{-1/2}$ (as defined in equation
\ref{epsl}) is shown versus $I$-band central surface brightness for
the sample of spiral galaxies of Mathewson \& Ford (1996) (with crude
bulge correction, see text).  On this plot, a fixed value of $\epsm$
corresponds to a horizontal line with amplitude $\epsm(\m2l
h^{-1})^{1/2}$.  The horizontal solid line shows the value of $\epsm$
for a self-gravitating disk ($\epsm=0.63$) with $\m2l=2.14h$. The two
shaded lines bracket the predicted ranges (see text).  Barred galaxies
are shown as triangles. }
\label{datfig}
\end{figure}

\section{Comparison with observations}\label{obsec}
To compare a theoretical instability criterion with observations, we
need both detailed rotation curves (to obtain $\vmax$) and surface
photometry (to obtain $\rd$).  We have examined the data from a wide
variety of sources.  The largest data set is that of Mathewson \& Ford
(1996) (MF) which has rotation velocities and $I$-band photometry for
a sample of nearly 2500 Southern spiral galaxies taken from the
ESO-Uppsala catalogue.  The majority are relatively late types: of
those 2275 for which Hubble types are given, 1055 are Sc or Sbc; 814
are Sb; and 5 are Sa.  We convert the published photometric quantities
to $\rd$ and the central surface brightness $\mu_0$ by assuming an
exponential profile.  Details are given in an Appendix.  The Appendix
also describes how we make a crude bulge subtraction from the MF data,
the results being labelled MFD in Table \ref{dattab}.  Since Mathewson
\& Ford (1996) provide the largest uniform data set we adopt it
whenever we quote a numerical result (including bulge subtraction when
not explicitly specified).  We list the other data sets for comparison
purposes.

\begin{table}
 \centering

  \caption{Observational data and summary of limits on $\m2l$.  Column
  (1), Name of data set; (2), number of galaxies; (3) median central
  surface brightness $\langle\mu_0\rangle$ in mag/sec$^2$; (4),
  $\epsl$ minimum, maximum, median and 10\% quantile values; (5)
  $\m2l$ limit derived from 10\% quantile and assuming $\epsm>0.63$;
  (6) assuming $\epsm>0.72$. See text for description of data set
  names. The row labelled `All' contains the combined properties of
  all the preceding rows.  MFB refers to the subset of MF which is
  barred.}

  \begin{tabular}{@{}lllrll@{}}
   Set & $N$   &$\langle\mu_0\rangle$& 
   $\epsl$[min,max,med,10\%] & $\m2l$(1)
   & $\m2l$(2) \\
   MF       &2446 &$19.97$& $[0.43, 7.50, 1.19, 0.92]$ & $ 2.14$ & $ 1.64$ \\
   C97      & 304 &$20.14$& $[0.54, 2.03, 1.13, 0.90]$ & $ 2.02$ & $ 1.55$ \\
   C96      & 316 &$20.39$& $[0.76, 1.84, 1.15, 0.96]$ & $ 2.34$ & $ 1.79$ \\
   SML      &  23 &$22.87$& $[0.54, 5.72, 1.95, 1.04]$ & $ 2.72$ & $ 2.08$ \\
   SMH      &  23 &$20.97$& $[0.67, 3.56, 1.32, 0.80]$ & $ 1.62$ & $ 1.24$ \\
   BBFN     & 371 &$18.47$& $[0.35, 2.10, 0.99, 0.70]$ & $ 1.24$ & $ 0.95$ \\
   DG       &  84 &$20.05$& $[0.59, 1.77, 1.04, 0.84]$ & $ 1.79$ & $ 1.37$ \\
   RBKG     &  25 &$20.89$& $[0.89, 1.78, 1.28, 1.08]$ & $ 2.93$ & $ 2.24$ \\
   B        &  12 &$19.89$& $[0.67, 1.96, 1.12, 0.84]$ & $ 1.77$ & $ 1.35$ \\
   \hline
   All      &3604 &$20.01$& $[0.35, 7.50, 1.16, 0.88]$ & $ 1.96$ & $ 1.50$ \\
   \hline
   MFD      &2446 &$19.97$& $[0.45, 7.50, 1.30, 0.98]$ & $ 2.41$ & $ 1.85$ \\
   MFB      & 175 &$19.94$& $[0.54, 4.37, 1.20, 0.90]$ & $ 2.05$ & $ 1.57$ \\
\end{tabular}
\label{dattab}
\end{table}

The data of Courteau (1996) and Courteau (1997) are in principle the
same, but one paper gives kinematic measurements and the other
photometric, and we list them separately.  To obtain the missing data
in these cases we use the observed Tully-Fisher relation (Giovanelli
et al. 1997; Shanks 1997):
\begin{equation} 
\MI - 5\log h = -(21.00 \pm 0.02) - (7.68\pm 0.13) (\log W - 2.5),
\label{TF}
\end{equation}
where $h$ is the Hubble constant in units of $100~\kms \mpc$,
and where $W$ is the inclination-corrected width of the HI line
profile.  The maximum of the rotation curve $\vmax$ is given by
$\vmax=W/2$.  Courteau (1997) has measured rotation not in HI, but in
H$_\alpha$.  Raychaudhury \etal{.} (1997) show that to a good
approximation values of $W$ determined from H$_\alpha$ are consistent
with those determined from HI.

For data sets without $I$-band quantities we convert using for all
galaxies $B-I=1.7$ (de Jong 1996) and $r-I=0.77$ (Rix \& Courteau 1997).
When $\rd$ or $\mu_0$ are missing we calculate them assuming an
exponential profile.

The other data sources are as follows with a short description of the
quantities provided.  C97: ($\rd$, $\vmax$) provided by the author from
Courteau (1997). C96: Courteau (1996) ($\rd$, $r$-magnitudes).  SML:
de Blok \& McGaugh (1996) low surface brightness galaxies ($\rd$,
$\vmax$, $B$-magnitudes). SMH: de Blok \& McGaugh (1996) high surface
brightness galaxies (ditto). BBFN: Burstein \etal{} (1996) spiral
types ($\vmax$, effective $B$-band photometric quantities). DG: Dale
\etal{} (1995), ($\rd$, $\vmax$, $I$-magnitudes and $\mu_0$). RBKG:
Raychaudhury \etal{} (1997), ($\rd$, $\vmax$, $I$-magnitudes). B:
Bottema (1993), ($\rd$, $\vmax$, $B$-magnitudes).

The luminosity (in solar units) is derived from listed magnitudes
using $L_I=10^{0.4(4.15-{\cal M}_I)}$.  We can then directly compute
the quantity
\begin{equation} 
\epsl = {\vmax \over (GL_I/\rd)^{1/2}},
\label{epsl}
\end{equation}
which is related to $\epsm$ by 
\begin{equation}
\epsl=\epsm\m2l^{1/2}.
\label{epslm}
\end{equation}
The quantity $\epsl^2$ has the units of a mass-to-light ratio, and
indeed it is a measure of the {\em total} mass (including dark matter
halo) contributing to the rotation curve.  Let us define a quantity
\begin{equation} 
\m2l^{\rm tot}(R) = {v^2|{c}(R) R  \over GL_I(R)},
\label{upstot}
\end{equation}
which measures the total mass-to-light ratio as a function of radius.
For an isolated disk it is a constant ($=\m2l$) and with an extended
dark halo it increases with radius.  The maximum rotation velocity in
general occurs at $R=R|{m}>\rd$, and the luminosity enclosed is
${L_I}|{,m}<L_I$, hence $\epsl^2<\m2l^{\rm tot}(R|{m})$.  
For an isolated disk
$R|{m}=2.2\rd$ and $L_{I,\rm m}=0.65L_I$, so
$\m2l=\Upsilon_I^{\rm tot}(R|{m})=3.4\epsl^2$.

In Figure \ref{datfig}, we show $\epsl$ as a function of central
surface density $\mu_0$ for the MFD data. The extra factor of $h$ in
the abscissa makes the plotted quantities independent of the Hubble
constant.  The figure reveals a marked correlation between $\epsl$ and
$\mu_0$.  This is expected if $\m2l$ is independent of $\mu_0$ in the
standard picture of disk formation (Fall \& Efstathiou 1980).  The
trend can be understood with a simple model where the dark matter halo
is assumed to be a singular isothermal sphere and disk self-gravity
is ignored (Mo, Mao \& White 1997). In this model, the disk scale
length and surface mass density scale as
\begin{equation} 
\rd \propto \lambda \rh,~~
\Sigma_0 \propto {M_d \over 2\pi \rd^2} \propto
{m_d v|m \over \lambda^2} H(z),
\label{rd-sigma}
\end{equation}
where $m_d$ is the fraction of halo mass that settles into the disk,
$\lambda$ is
the dimensionless spin parameter, $v|m$ is the halo circular velocity
and $H(z)$ is the Hubble constant at redshift $z$ (see Mo, Mao \& White
1997 for details). The Tully-Fisher relation is given by
\begin{equation} 
L_I = A v^3|m, ~~ A \equiv f_l
{m_d\over \m2l h^{-1}}{H_0\over H(z)},
\label{tf}
\end{equation}
where $f_l$ is a dimensionless constant which depends on the cosmology
and halo profile (Mo, Mao \& White 1997).
From equations (\ref{rd-sigma}) \& (\ref{tf}), we have
\begin{equation} 
\epsl h^{-1/2} \propto \lambda^{1/2} A^{-1/2} 
\left({H_0 \over H(z)}\right)^{1/2},
\label{epslA}
\end{equation}
and
\begin{equation}
\mu_0 \propto {v|m \over \lambda^2}A \left({H_0 \over H(z)}\right)^{-2},
\label{IlA}
\end{equation}
where $\mu_0\equiv \Sigma_0/\m2l$.
Notice that $m_d, \m2l$ and the Hubble constant 
come into the above expressions only through 
$A$ defined in equation (\ref{tf}). From equations (\ref{epslA}) and
(\ref{IlA}) we obtain the dependence
of $\epsl$ on $\mu_0$ as
\begin{equation} \label{epsl-mu}
\epsl h^{-1/2}\propto \mu_0^{-1/4} v^{1/4}|m A^{-1/4}.
\label{eps-pred}
\end{equation}
For a given central surface brightness, the scatter in $\epsl$ is
determined by the range in circular velocities and the scatter in the
Tully-Fisher amplitude $A$. Since the scatter in $A$ is approximately
a factor of 2 and the velocity range is $\approx 100-300~\kms$, we
expect a scatter of roughly 60 percent. In Figure \ref{datfig}, we
overlay the predicted slope and scatter on top of the data points with
the normalization chosen to reproduce the observed median value of
$\epsl$ in the range $\mu_0\in(20,21)$.  The lines are derived from
the 25 and 75\% quantiles of $A/\vmax$ of the data---the equivalent
scatter in $A$ is a factor of $\approx 1.9$ or $0.7$ magnitudes.  As
can be seen, the predicted slope and scatter are consistent with the
observed data points. Thus, the data are consistent with the
assumption that the value of $A$ is independent of $\mu_0$ for these
galaxies.  At the fainter end the observed $\epsl$ is slightly higher
than the model prediction, implying either $\m2l$ is higher or
$m_d$ is lower for low-surface-brightness galaxies.  The former would
result from a lower star formation efficiency, as suggested by
observations (e.g. McGaugh \& de Blok 1997). Since $\epsilon_l \propto
\lambda^{1/2}$ (cf. equation \ref{epslA}), a substantial angular
momentum loss would lower $\epsilon_l$ significantly and make many
disks unstable. Hence, the stability of disks require approximate
conservation of angular momentum in the disk formation process.  The
normalization in equation (\ref{epsl-mu}) is dependent on the detailed
halo profiles and other factors.  A detailed comparison with the
observations would also need to take into account observational biases
(\eg{,} in surface brightness), which is beyond the scope of the
present work.

Assuming that the disks are exponential already sets a limit on
$\m2l$: they should all have $\epsm>0.63$, the value for an isolated
disk.  To be conservative, let us suppose that 90\% of galaxies should
have $\epsm>0.63$, and ascribe the remaining 10\% to deviations from
exponential and/or measurement errors.  In Figure \ref{datfig} the
solid horizontal line marks the 10\% quantile of $\epsl$ in the data
of Mathewson \& Ford (1996) (after crude bulge subtraction, see
Appendix).  This corresponds to $\epsm=0.63$ for $\m2l=2.41h$.  When
the bulge subtraction is not carried out the implied value of $\m2l$
goes down by about 20 percent to $\m2l=2.14h$.  The quantiles and
implied limits on $\m2l$ for the other data sets are given in Table
\ref{dattab}.  The differences between the data sets are on the whole
not large.  Burstein \etal{} (1996) (BBFN) give only effective
quantities, and hence we believe this data is more heavily
contaminated with bulge contributions than the others.  This would
account for the lower values of $\epsl$ and $\m2l$ (and also for the
brighter median $\mu_0$).

If $\m2l>2.41h$ more than 10\% of galaxies would have $\epsm<0.63$
which would be incompatible with their having exponential surface
density distributions.  Thus a conservative upper limit on the
mass-to-light ratio of disks is $\m2l<2.41h$.  Note that this limit
applies only under the assumption that $\m2l$ is independent of
$\mu_0$. This assumption is, as we argued, consistent with the
observational data, if we adopt the standard picture of disk
formation.  The derived limit is also consistent with independent
measurements of $\m2l$ (see Section 6 for details).  However, since
$\epsl$ is higher for low-surface-brightness galaxies, a higher upper
limit on $\m2l$ is still allowed for these galaxies without violating
the constraint $\epsm>0.63$.

A tighter upper limit on $\m2l$ is provided by demanding that disks
are dynamically stable.  The stability criterion proposed by \ELN{}
($\epsm<1.1$) implies a very low value of $\m2l$: the 10\% quantile
giving an upper limit of $\m2l<0.79h$ for the MFD data.  This upper
limit is very low and is inconsistent with direct measurements of
$\m2l$.  To try and resolve this conflict we carried out our own
$N$-body simulations to locate the stability boundary independently.

\section{Numerical simulations}\label{numsec}
Here we describe $N$-body simulations of exponential disks embedded in
dark halos.  The halos we use, in common with \ELN{}, are rigid
(\ie{.} they are modelled by a fixed potential).  The halo density
profile used was either a Hernquist (1990) profile, or a truncated
isothermal sphere as described in Hernquist (1993).  A comparison
between the two halo types probes the effect of halo concentration on
disk stability.  The disk is locally isothermal:
\begin{equation}
\rho(R,z) = {\Sigma(R)\over2z_0} {\rm sech}^2(z/z_0)
\label{isoth}
\end{equation}
with a constant
vertical scale height of $z_0=0.2\rd$ at all radii.  The radial
dispersion velocity is given by 
\begin{equation}
\langle v_R^2\rangle = C \exp(-R/\rd)
\label{vReq}
\end{equation}
with $C$ constant.  This is consistent with observations of galactic
disks (van der Kruit \& Searle 1981, Bottema 1987).  The constant $C$
is chosen by fixing $Q=1.2$ (Toomre 1963) at $R=2.4\rd$ (this is then
roughly the minimum $Q$ in the disk).  Galactic stellar disks may well
have larger values of $Q$ (Lewis \& Freeman 1989), but this would lead
to greater stability, and hence the limit we derive will be
conservative.  The disk equilibria were set up as described by
Hernquist (1993) using a code provided by Professor Hernquist himself.

\begin{figure}
\epsfxsize=.8\hsize
\computelabs
\setlabel{v|c\over\sqrt{GM|d/\rd}}{-.01}{-.5}
\setlabel{R/\rd}{.6}{-0.95}
\centerline{\epsfnormal{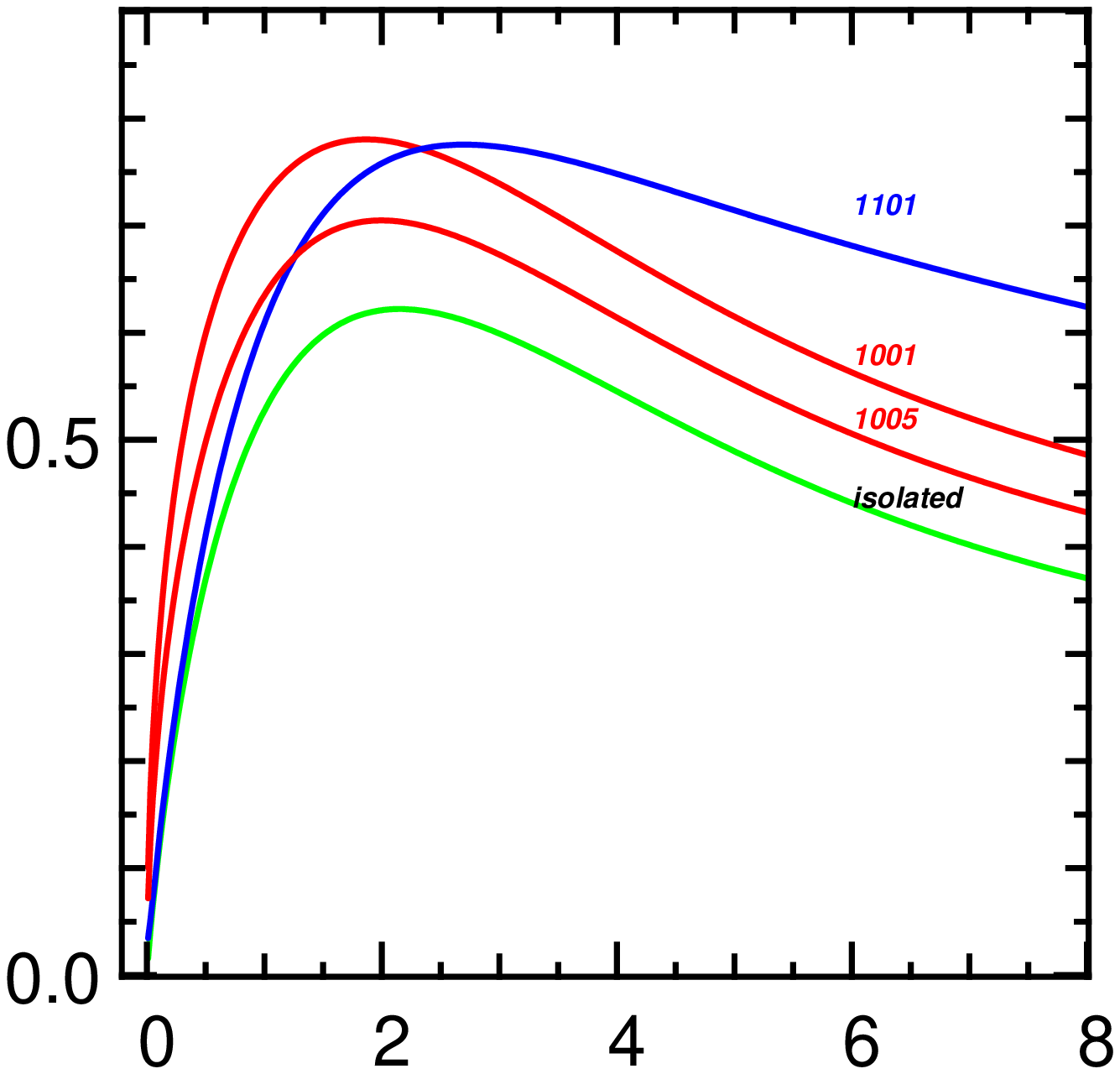}}
\vskip\baselineskip
\caption{ The rotation curves of some of the models in Table
\ref{simtab}.  The lower curve is that of an isolated
exponential disk. 
}
\label{vcfig}
\end{figure}

We used the AP$^3$M code of Couchman, Pearce and Thomas (1995)
compiled for isolated (not cosmological) initial conditions, and
modified to include the rigid halo contribution.  The simulations had
16384 disk particles, and a softening length of $0.05\rd$.  Each
simulation was run for approximately 7 times
$2\pi(\rd^3/GM|d)^{1/2}$. Table \ref{simtab} lists the simulation
parameters in full.  The rotation curves of some of the models are
shown in Figure \ref{vcfig}.  The Hernquist halos have cuspy profiles
and hence the rotation curves rise much more steeply than in the
isothermal case for the same scale length.  The Hernquist rotation
curves are also more peaked than the isothermal case.  Peakiness is
not in this instance an indication of low $\epsm$ but rather of the
finite mass of the halo: the rotation curves are roughly Keplerian for
$R\ga3$.

\begin{table}\label{simtab}
 \centering

  \caption{Simulation parameters.  Lengths in units of $\rd$; masses
  in units of $M|d$.  LH$=$Hernquist profile; IS$=$isothermal.  $M|h$
  is total halo mass.  The IS halos were truncated exponentially at
  $r=8$ (Hernquist 1993); $a$ is halo scale length; $\epsm$ is
  defined in equation (\ref{epsdef})}

  \begin{tabular}{@{}llllll@{}}
   Name     & Halo   & $M|h$ & $a$  & $\epsm$ & bar\\
   1001     & LH     & $1$     & 1      & $0.78$  & no      \\
   1002     & LH     & $2$     & 1      & $0.92$  & no      \\
   1004     & LH     & $4$     & 1      & $1.15$  & no      \\
   1005     & LH     & $.5$    & 1      & $0.70$  & yes      \\
   1101     & IS     & $2.5$   & 1      & $0.78$  & no      \\
   1102     & IS     & $5$     & 1      & $0.92$  & no      \\
   1104     & IS     & $10$    & 1      & $1.16$  & no      \\
   1103     & IS     & $2$     & 1      & $0.75$  & no      \\
   1105     & IS     & $1.25$  & 1      & $0.70$  & yes      \\
\end{tabular}
\end{table}

We can evaluate the strength of the instability visually and
quantitatively.  The endpoints of simulations 1005 and 1001 are shown
in Figure \ref{barfig}, and a bar is clearly visible in the
former.  As an objective measure we use a logarithmic spiral
decomposition of the surface density (Anathassoula \& Sellwood 1986).
We define the quantity
\begin{equation}
A_p = {\sqrt{N}\over M|d}\; \sum\limits_{j=1}^{N} m_j \exp[i(2\theta_j + p\ln
R_j)], 
\label{apeq}
\end{equation}
where $p$ is the logarithmic spiral wavenumber, $m_j$ is the mass, and
$(R_j,\theta_j)$ are the cylindrical co-ordinates of a particle
labelled by an index $j$.  The factor outside the summation scales
$A_p$ so that the expected value is unity if there is no significant
structure at a given $p$.  The factor of 2 in the argument of the
exponential means that we pick out density perturbations with $m=2$
symmetry.  For $p>0$ we have a trailing spiral and for $p<0$ a leading
spiral.  A bar has $p=0$.  We measure $A_p$ approximately 40 times
during each simulation, and consider the quantity $\mod{A_0}$ defined
as the average over the range $p\in[-1,1]$.  In Figure \ref{Afig} we
show $\mod{A_0}$ as a function of time for all the simulations.  Those
with visually identifiable bars (1005, 1105) show up clearly as having
a significantly larger value of $\mod{A_0}$.  Thus both methods agree.
We give an indication of whether a bar was present or not in Table
\ref{simtab}.

\begin{figure}
\epsfxsize=.95\hsize
\centerline{\epsfnormal{barfig.ps}}
\caption{ The endpoints of two of the simulations (showing only 4096
particles to avoid crowding).  Simulation 1005 clearly shows a strong
bar, while 1001 does not. The rotation curves of these two models are
shown in Figure \ref{vcfig}.  The figures are bounded by the
simulation box which is 8$\rd$ on a side.
}
\label{barfig}
\end{figure}

\begin{figure}
\epsfxsize=.8\hsize
\computelabs
\setlabel{\mod{A_0}}{.1}{-.5}
\setlabel{t/T|d}{.64}{-.93}
\centerline{\epsfnormal{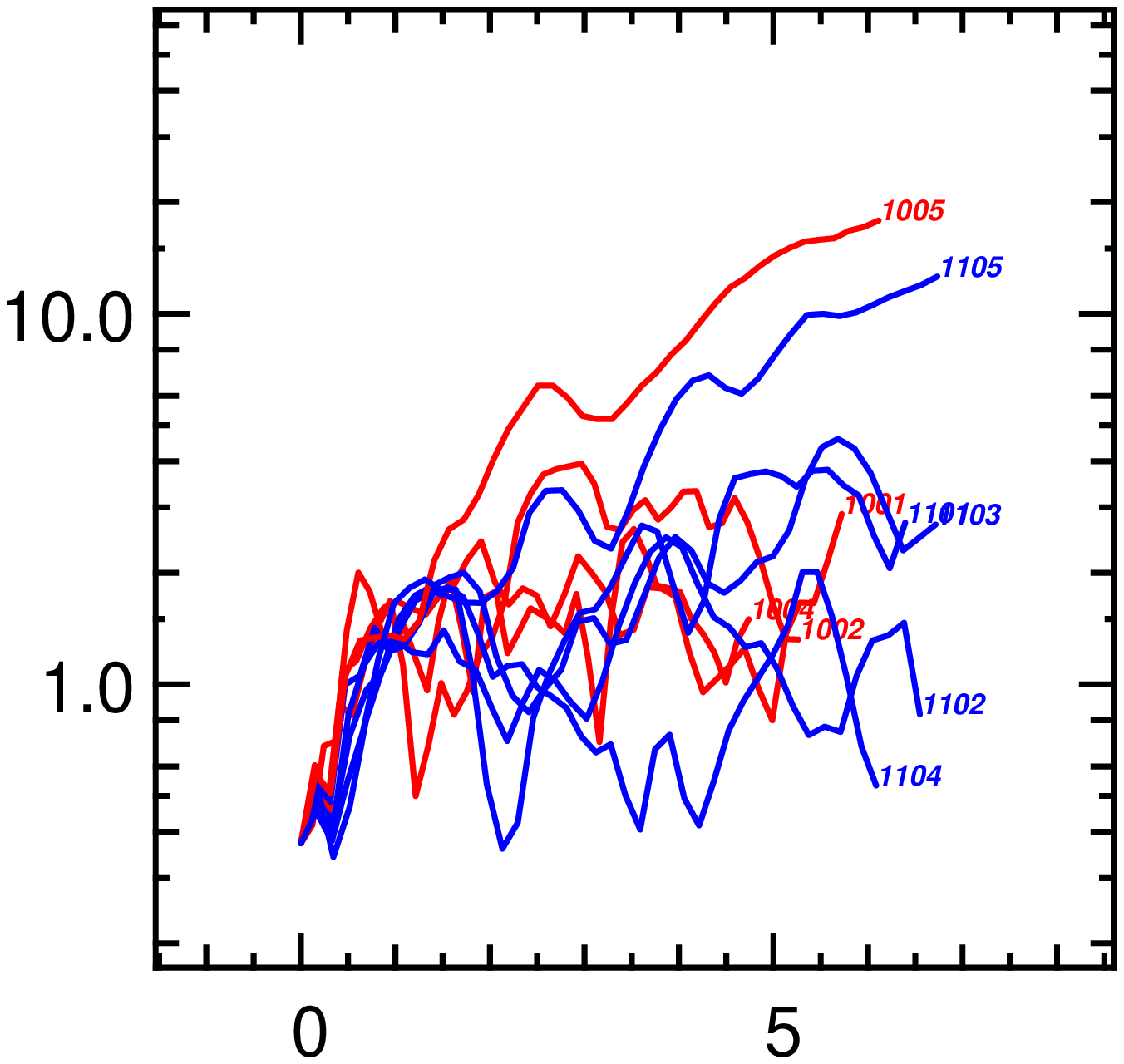}}
\vskip\baselineskip
\caption{ The strength of the bar $\mod{A_0}$ as a function of time
for the simulations in Table \ref{simtab}.  Simulations 1005 and 1105
show clearly the presence of a bar, as shown in Figure \ref{barfig}.
The time unit is $T|d=2\pi(\rd^3/GM|d)^{1/2}$.
}
\label{Afig}
\end{figure}

It follows that the boundary $\epscr$ between global stability and a
bar-forming instability lies between $\epsm=0.70$ and $\epsm=0.75$.
The 10\% quantile in Figure \ref{datfig} combined with these values of
$\epscr$ gives an upper limit to the mass-to-light ratio
$\m2l<(1.85\pm.1)h$.  The formal error only reflects the uncertainty in
the value of $\epscr$.  The upper limit would be higher if one had
more tolerance for disks with apparent $\epsm<\epscr$.  Nevertheless
the limit is in agreement with independent direct measurements (see
Section \ref{detsec}).

The critical value $\epscr$ is much lower than that obtained by \ELN{}
for their `FE' models ($\epscr\approx1.1$).  \ELN{} also conducted
simulations with isothermal halos similar to ours, but they did not
cover as much parameter space as their FE models, and were not
analyzed in much detail.  The obvious differences between our
simulations and those of \ELN{} are that our simulations are
3-dimensional, and that they have slightly better formal resolution.
Resolution would not seem to explain our different conclusion.
Indeed, we carried out simulations with $N=1024$ and a significantly
larger softening length and reached the same conclusion.  The
difference could arise from the finite thickness of our disks.  They
have a vertical scale height of $0.2\rd$, implying a formal
eccentricity of $e=\sqrt{1-.2^2}=0.97$.  Isolated stellar Maclaurin
spheroids are already unstable at $e=0.97$.  The added external
potential may be enough to stabilize our disks but not those of \ELN{}
with $e=1$.  A further difference between our simulations and those of
\ELN{} is the value of Toomre's $Q$ and its dependence on radius.
Most of the simulations of \ELN{} use $Q(R)=\const=1.05$, whereas we
use a profile with a broad minimum of $Q=1.2$ at $R\approx2.4\rd$.
\ELN{} also ran a few simulations with $Q$ a decreasing function of
radius, and reported that their results were unchanged.  The lower
value of $Q$ would make their disks more prone to bar formation.  Note
however that typical values for galactic disks are even higher than in
our simulations (Lewis \& Freeman 1989, Bottema 1987, von Linden \&
Fuchs 1997).

Our results support the conclusion of \ELN{} that $\epsm$ is a good
indication of stability, independent of halo type and concentration
(the Hernquist halos are more concentrated than the isothermal ones).
One implication is that a modified Ostriker \& Peebles criterion such
as that considered by Christodoulou, Schlosman \& Tohline (1995) is
unlikely to be universal.

\section{Independent determinations of $\Upsilon$}\label{detsec}
Limits on the mass-to-light ratio of the Galactic disk in the solar
neighbourhood can be derived from a combination of kinematic
measurements and star counts.  Kuijken \& Gilmore (1989) derive a
local surface mass density in the disk of $40\Msun/\pc^2$, and star
counts give a $V$-band luminosity density of $15\Lsun/\pc^2$ (Gould,
Bahcall \& Flynn 1996).  Dividing mass by light we obtain
$\Upsilon_V=2.67$, and thus $\m2l\approx 1.9$ (assuming $V-I=1.0$).
This number is independent of $h$, and hence is rather high compared
with the upper limit derived from the $N$-body simulations.  The data
quoted by Bottema (1993) give a value of $\epsl=1.37$ for the Galaxy,
so for $\m2l \approx 1.9$ we obtain $\epsm\approx 1.0$.  Thus despite
having high $\m2l$ we conclude that the disk of the Galaxy should be
globally stable.

Mass-to-light ratios can also be derived from pure stellar population
synthesis arguments, although there is always some uncertainty arising
from the poorly known initial mass function (IMF), particularly from
the low-mass cut-off in the IMF. A few stellar population models are
available (e.g., Bertelli et al 1994; Bruzual \& Charlot 1993; Worthey
1994).  For a Salpeter IMF, the mass-to-light ratio derived by various
authors appear to agree within an accuracy of 25\% (Charlot, Worthey
\& Bressan 1996). The predicted $\m2l$ depends on the metallicity and
age of the stellar population. For a stellar population with age
between 5-12 Gyr and with a solar metallicity, $\m2l$ is between
$0.9$-$1.8$ for a constant star formation rate (\cf{} Table 3 in de
Jong 1996).  For an exponential star formation law, the mass-to-light
ratio is about 20\% higher. The predicted values are in good agreement
with the values derived from the instability analysis presented in
this paper. In the comparison, we have neglected the uncertainty due
to dust, since the Tully-Fisher studies already attempt to correct for
its effect.  Furthermore de Jong (1996) argues that dust reddening
probably plays a minor role in the color gradients in disk
galaxies. Nevertheless, the dust correction remains a nuisance in
these comparisons.

\begin{figure}
\epsfxsize=.8\hsize
\computelabs
\setlabel{\m2l/h}{-.0}{-.5}
\setlabel{\mu_0 (I-\hbox{mag/sec}^2)}{.55}{-.94}
\centerline{\epsfnormal{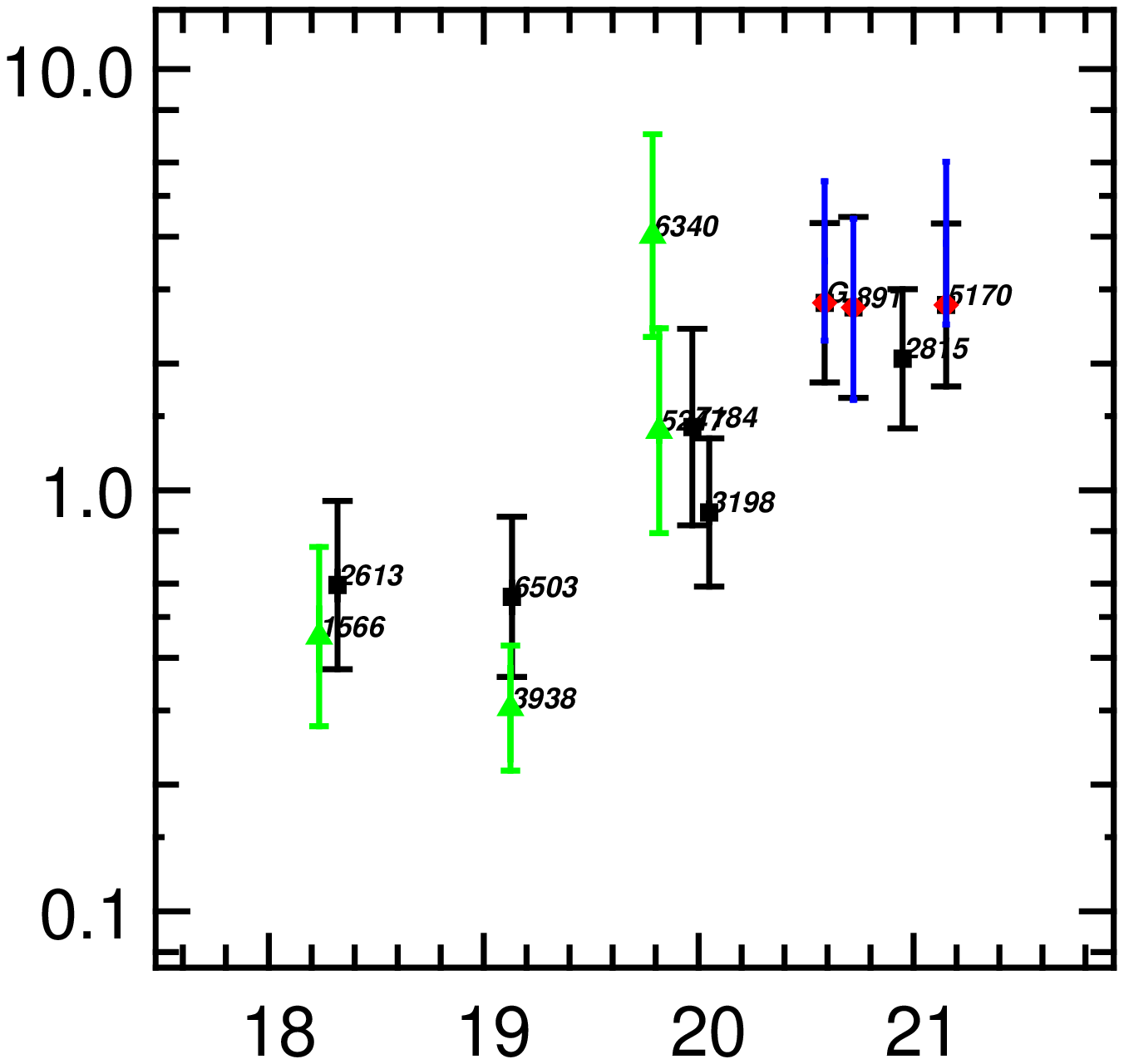}}
\vskip\baselineskip
\caption{ The mass-to-light ratio versus I-band central surface
brightness for 12 galaxies from Bottema (1993)
calculated according to equation (\ref{upseq}).  Each point is marked by
its NGC number (or G for the Galaxy).  For the Galaxy $\m2l$ has not
been divided by $h$.  Squares are for inclined galaxies; triangles for
face-on galaxies; circles are additional points for highly inclined
galaxies where $z_0$ is measured (otherwise $z_0/\rd=0.2$ is assumed).
The errorbars are estimated from the errors in the velocity
dispersions.
}
\label{botfig}
\end{figure}

A direct measurement of the mass-to-light ratio of extragalactic disks
requires detailed kinematic studies such as that described by Bottema
(1993).  Bottema (1997) derives a value of $\m2l=(1.7\pm0.5)h$ after
somewhat uncertain processing of his data. 
The measurement of $\Upsilon$ relies on
the relation between vertical velocity dispersion $\sigma_z$, surface
density $\Sigma$ and vertical scale height $z_0$.  Essentially, the
larger the value of $\Sigma$, the hotter a disk has to be at constant
$z_0$.  For a disk which is well approximated by an isothermal sheet
(cf. equation \ref{isoth})
\begin{equation}
\sigma_z^2 = \pi G \Sigma z_0
\label{sigeq}
\end{equation}
(\eg{.} Binney \& Tremaine 1987).  Assuming an exponential radial
profile we can write the mass-to-light ratio as
\begin{equation}
\Upsilon = {2G\sigma_z^2 \rd\over L|d} {\rd\over z_0}.
\label{upseq}
\end{equation}
Taking account of the possible variations in vertical density profile
in the disk, equation (\ref{upseq}) represents an upper limit on
$\Upsilon$ (Wielen \& Fuchs 1983).  Bottema measures
$\sigma_z(R\!=\!0)$ for face-on galaxies, and asserts that
$\sigma_R(R\!=\!\rd) = \sigma_z(R\!=\!0)$ for inclined galaxies (this
is true for the Galaxy).  Only 4 of the 12 galaxies are edge-on enough
to measure $z_0$, so an assumption also has to be made about its value
in the other galaxies.  Bottema discusses this problem in some depth.
Where it can be measured directly $z_0/\rd\approx0.2$ and does not
vary by more than about 20 percent (Barteldrees \& Dettmar 1994).

Figure \ref{botfig} shows $\m2l$ calculated from equation
(\ref{upseq}) (assuming $z_0/\rd=0.2$ where no measurement is
available) plotted against $\mu_0$.  The quantity plotted is
independent of $h$. In the case of the Galaxy $\m2l$ has not been
divided by $h$.  The first impression of this figure is a marked trend
of smaller $\m2l$ for higher surface brightness galaxies (low $\mu_0$
in magnitudes).  The range of values of $\m2l$ is also large (about a
factor of $10$ between smallest and largest values).  This is in
marked conflict with the received wisdom regarding mass-to-light
ratios of galactic disks, based on the uniformity of the colours of
stellar populations (de Jong 1996).  Indeed taken at face value, the
trend in Figure \ref{botfig} is so strong that when applied to 
equation (\ref{epsl-mu}) it should produce an {\em anti-}correlation 
between $\epsl$ and $\mu_0$ for constant $m_d$.  


\begin{figure}
\epsfxsize=.8\hsize
\computelabs
\setlabel{\epsm}{0.05}{-.5}
\setlabel{\mu_0 (I-\hbox{mag/sec}^2)}{.55}{-.94}
\centerline{\epsfnormal{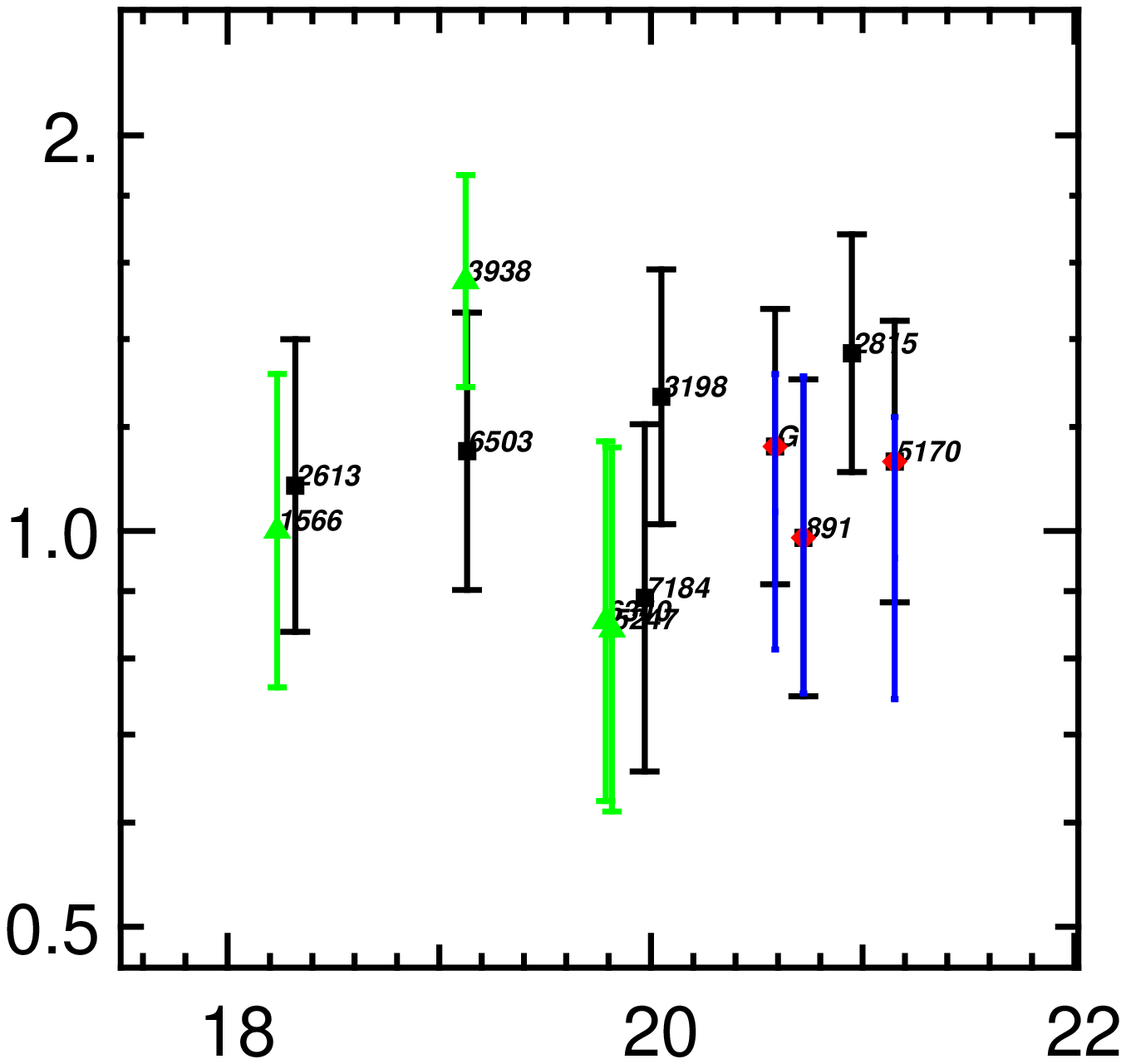}}
\vskip\baselineskip
\caption{ Shows $\epsm$ for the galaxies of Bottema (1993) given
the values of $\m2l$ from Figure \ref{botfig}.
}
\label{botem}
\end{figure}

How should we interpret Figure \ref{botfig}?  We would like to see
many more data before any strong conclusions are drawn.  One can
separate the galaxies by eye into two groups: one with high surface
brightness ($\mu_0<19.5$) and low $\m2l$, and the other with lower
surface brightness ($\mu_0>19.5$) and higher $\m2l$.  Of the four
galaxies with $\mu_0<19.5$, two have significant non-stellar
contributions to their luminosity (3938 and 1566) which could bias
them artificially towards the bottom left of Figure \ref{botfig}.
Notice that few galaxies in the MF sample have surface brightness
as high as these four galaxies.  Taken on their own, the galaxies with
$\mu_0>19.5$ have values of $\m2l$ which are perhaps not a strong
function of $\mu_0$.

From equations (\ref{epsdef}) \& (\ref{upseq}) we can obtain a simple
expression for $\epsm$ in terms of kinematical quantities:
\begin{equation} 
\epsm = {\vmax \over \sigma_z} \sqrt{z_0\over 2\rd}.
\label{epsmv}
\end{equation}
Figure \ref{botem} shows the values of $\epsm$ for the galaxies in
Figure \ref{botfig} calculated from equation (\ref{epsmv}).  They are
all $\ga0.75$ so the values of $\m2l$ in Figure \ref{botfig} are
consistent with the disks being globally stable.

There remains a constraint on the average $\m2l$ as a function of
$\mu_0$ from the theory of disk formation.  As was mentioned in
Section \ref{obsec}, the observed positive correlation in Figure
\ref{datfig} is consistent with $A$ being at most a weak function of
$\mu_0$, particularly for the high surface brightness galaxies. This
conclusion is expected to be only weakly dependent on the detailed
model of disk formation---only the normalisation depends somewhat on
the cosmological model (Mo, Mao \& White 1997). The mass-to-light
ratio $\m2l$ required to match the Tully-Fisher relation is in the
range of 1 to $2h$, consistent with those obtained from the
stability analysis presented here and models of stellar population
synthesis. Whether it is actually consistent with Figure \ref{botfig}
is unclear until we have more data. If such a trend exists, the small
scatter in the Tully-Fisher amplitude $A$ would require a nearly
linear correlation between $m_d$ and $\m2l$ (\cf{} equation \ref{tf}).

\section{Discussion}\label{dissec}
Global instability in a disk leads to bar formation, so it is natural
to assume that barred galaxies form in this way (\eg{.} Sellwood
1996).  If the correlation of $\epsl$ with $\mu_0$ really reflects a
correlation of $\epsm$ with $\mu_0$, then we should expect galaxies
with smaller values of $\epsl$ to form bars.  No such effect is
visible in the MF data.  The distribution of barred types is
indistinguishable from that of disk galaxies as a whole (Figure
\ref{datfig}).  Bar formation does not have to be spontaneous: it
could be induced by a perturbation from a close encounter with another
galaxy (Noguchi 1996), or from an interaction between the disk and the
halo.  Perhaps all disks have $\m2l$ such that a bar does not form
spontaneously, but they are equally susceptible to induced bar
formation.  This is consistent with the values of
$\epsm$ in Figure \ref{botem}.  Note however that the barred fraction
in the MF data is low ($\approx10\%$) compared to that reported by other
authors (\eg{.} Sellwood \& Wilkinson 1993 give a fraction of around
30\%).  Further discussion of this important question is outside the
scope of the present work.

Clearly it is important to extend the range of data analysed in
Section \ref{detsec}.  This requires a good HI rotation curve for each
galaxy, and high quality spectroscopy at least along the major axis of
the galaxy.  Ideally one would choose the brightest members of a large
pre-defined sample (such as that of Mathewson \& Ford 1996) and follow
them up with a high spatial resolution spectrograph.  Kinematic
information in more than one dimension is also advantageous since it
removes some of the uncertainties in the deprojection of the velocity
ellipsoid.  A number of two dimensional spectrographs are due to come
on line shortly, and these may be well suited to the problem.  The
biggest uncertainty in the determination of $\m2l$ will remain that
associated with the value of $z_0$.  Efforts should therefore be made
to analyse as many edge on galaxies as possible to try to improve on
existing determinations of the distribution of $z_0/\rd$ (\eg{.}
Barteldrees \& Dettmar 1994).

\section*{Acknowledgments}

We are grateful to Simon White for helpful discussions.  This project is
partly supported by the ``Sonderforschungsbereich 375-95 f\"ur
Astro-Teilchenphysik'' of the Deutsche Forschungsgemeinschaft.

{}

\section*{Appendix}
Here we describe how values of $\rd$, $\mu_0$ and $L$ were derived
from the published quantities of Mathewson \& Ford (1996).  The
published data list total magnitudes, and face-on corrected isophotal
quantities: average surface brightness and isophotal diameter at
$\mu=23.5 \hbox{mag/sec}^2$.  Assuming an exponential disk we have a
surface brightness profile
\begin{equation}
\mu(R) = \mu_0 \exp(-\alpha),~~\alpha \equiv {R \over R_d},
\label{mur}
\end{equation}
and for a bulge to disk ratio $f$ we have (outside the bulge)
\begin{equation}
L(R) = L|d \lr{1+f - \exp(-\alpha)(\alpha+1)}.
\label{lreq}
\end{equation}
 From equation (\ref{lreq}) the average surface
brightness inside radius $R$ is
\begin{equation}
\bar\mu = {L(R)\over\pi R^2} = {2\mu_0\over\alpha^2} 
\lr{1+f-\exp(-\alpha)(\alpha+1)},
\label{Iave}
\end{equation}
where we have used $L|d=2\pi\mu_0\rd^2$.  Combining equations
(\ref{lreq}) and (\ref{Iave}) we obtain
\begin{equation}
{\bar\mu\over\mu} = {\exp(\alpha)(1+f) - (\alpha+1)\over \alpha^2}. 
\label{mumeq}
\end{equation}
Given $f$ and $\bar\mu/\mu$ we can solve equation (\ref{mumeq})
numerically to find $\alpha$.  Then we have $\rd=R/\alpha$, and
$\mu_0=\mu\exp(\alpha)$.

In Table \ref{dattab} the data labelled MF is obtained by setting
$f=0$ (no bulge).  Those labelled MFD have a crude bulge subtraction
applied as follows.  If the galaxy is classified Sbc or Sc a value of
$f$ is initially chosen randomly from a uniform distrubution in the
range $[0,.2]$.  If it is Sa or Sb $f$ is similarly chosen in the
range $[0,.4]$.  This covers about 80\% of the galaxies, and for the
remainder we simply set $f=0$.  For too large a value of $f$, equation
(\ref{mumeq}) does not have a solution for positive real $\alpha$.  If
$f$ is initially too large, it is reduced by a factor of $3/4$.  This
factor is applied repeatedly until a solution is found.

\bsp
\label{lastpage}
\end{document}